\documentclass[traditabstract,times]{aa} 

\newcommand{\ergps}{erg\thinspace s$^{-1}$}
\newcommand{\phpspsqcm}{ph\thinspace cm$^{-2}$\thinspace s$^{-1}$}
\newcommand{\ergpspsqcm}{erg\thinspace s$^{-1}$\thinspace cm$^{-2}$}
\newcommand{\psqcm}{cm$^{-2}$}
\newcommand{\nH}{$N_{\rm H}$}

\usepackage{graphicx}
\begin{document}

\title{Highly ionized disc and transient outflows in the Seyfert galaxy IRAS~18325--5926}


   \author{K. Iwasawa\inst{1,2}\thanks{Email: kazushi.iwasawa@icc.ub.edu}
          \and
A.C. Fabian\inst{3}
\and
E. Kara\inst{3,4}
\and
C.S. Reynolds\inst{4}
\and
G. Miniutti\inst{5}
\and
F. Tombesi\inst{6,4}
}

\institute{Institut de Ci\`encies del Cosmos (ICCUB), Universitat de Barcelona (IEEC-UB), Mart\'i i Franqu\`es, 1, 08028 Barcelona, Spain
         \and
ICREA, Pg. Llu\'is Companys 23, 08010 Barcelona, Spain
\and
Institute of Astronomy, Madingley Road, Cambridge CB3 0HA, United Kingdom
\and
Department of Astronomy, University of Maryland, College Park, MD 20742-2421, USA
\and
Centro de Astrobiologia (CSIC-INTA), Dep. de Astrof\'isica, ESAC, PO Box 78, 28691 Villanueva de la Ca\~nada, Madrid, Spain
\and
X-ray Astrophysics Laboratory and CRESST, NASA/Goddard Space Flight Center, Greenbelt, MD 20771, USA
          }


 
          \abstract{We report on strong X-ray variability and the Fe K-band spectrum of the Seyfert galaxy IRAS 18325--5926
            obtained from the 2001 XMM-Newton EPIC pn observation with a duration of
            $\sim 120$ ks. While the X-ray source is highly
            variable, the 8-10 keV band shows larger variability than
            that of the lower energies. Amplified 8-10 keV flux
            variations are associated with two prominent flares of the
            X-ray source during the observation. The Fe K emission is
            peaked at 6.6 keV with moderate broadening. It is likely to
            originate from a highly ionized disc with an
            ionization parameter of log~$\xi\simeq 3$. The Fe K line
            flux responds to the main flare, which supports its disc
            origin. A short burst of the Fe line flux has no
            relation to the continuum brightness, for which we have no
            clear explanation. We also find transient, blueshifted Fe
            K absorption features that can be identified with
            high-velocity ($\sim 0.2\thinspace c$) outflows of highly
            ionized gas, as found in other active galaxies. The
            deepest absorption feature appears only briefly ($\sim 1$
            hr) at the onset of the main flare and disappears when
            the flare declines. The rapid evolution of the
            absorption spectrum makes this source peculiar among the
            active galaxies with high-velocity outflows. Another
            detection of the absorption feature also precedes the
            other flare. The variability of the absorption feature
            partly accounts for the excess variability in the 8-10 keV
            band where the absorption feature appears. Although no
            reverberation measurement is available, the black hole
            mass of $\sim 2\times 10^6M_{\odot}$ is inferred from the
            X-ray variability. When this mass is assumed, the black
            hole is accreting at around the Eddington limit, which may
            fit the highly ionized disc and strong outflows observed
            in this galaxy.}

\titlerunning{XMM-Newton observation of IRAS 18325--5926}
\authorrunning{K. Iwasawa et al.}
   \maketitle
%

\section{Introduction}

IRAS 18325--5926 (= Fairall 49) is one of the early X-ray selected
active galactic nuclei (AGN) of Picinotti et al. (1982), identified
with an obscured Seyfert nucleus hosted by the IRAS galaxy at
$z=0.0198$ (Ward et al. 1988; Carter 1984; de Grijp et al. 1985; Strauss
et al. 1992; Iwasawa et al. 1995). The nuclear X-ray source is
moderately absorbed by \nH $\approx 1\times 10^{22}$ \psqcm\ and shows
strong flux variability. The X-ray spectrum shows a
strong iron emission feature (Fe K). The line profile appears broad
and peaks at 6.6 keV, which Iwasawa et al. (1996, 2004) suggested to
originate from reflection of an highly ionized disc (but see Lobban \&
Vaughan 2013). This is unusual among nearby bright AGN in which a
cold line at 6.4 keV is the normal Fe K feature, suggesting that
this
active nucleus may be operating in an extreme condition.

XMM-Newton observations of IRAS 18325--5926 have been made in 2001 and 2013
with exposure times of 120 ks and 200 ks, respectively. In the recent
2013 observation, in which two 100 ks exposures are separated by three
weeks, Lobban, Alston \& Vaughan (2014) found hard X-ray lags. For the
2001 observation, data from the two EPIC MOS cameras, MOS1 and MOS2,
have been extensively analysed for the composition of the broad-band
X-ray spectrum and its variability by Tripathi et al. (2013) and Lobban
\& Vaughan (2013), while only the time-averaged spectrum was briefly
presented for the EPIC pn data in Iwasawa et al. (2004). Here we focus
on the Fe-K band spectrum and its variability during the 2001
observation, using the EPIC pn data, since the pn camera has a higher
sensitivity for the Fe K band and it is clearly advantageous
to study this
photon-limited band over the MOS cameras.

The cosmology adopted here is $H_0=70$ km s$^{-1}$ Mpc$^{-1}$,
$\Omega_{\Lambda}=0.72$, $\Omega_{\rm M}=0.28$.

\section{Observations}

We observed IRAS 18325--5926 with XMM-Newton in a full orbit of
revolution 227 on 2001 March 5-7 (ObsID 0022940301, PI: K. Iwasawa). The
EPIC pn camera was operating in small window mode with the
medium filter. As no disruptive background flares occurred during the
observation, we used the whole duration of the observation of
approximately 118 ks. With the efficiency of the small window mode
($\sim 70$ \%), the net exposure time is 81.5 ks.

The mean (net) count rate from the X-ray source in the 2-10 keV band
is $1.330\pm 0.004$ ct s$^{-1}$. The background count rate has a small
fraction ($\simeq 1$\%) of the source count rate in the same band.

\section{Results}

\subsection{Mean broad-band spectrum}

The broad-band (2-11 keV) X-ray spectrum obtained with the EPIC pn
camera is shown in Fig. 1. The spectrum is plotted in 200 eV
intervals. It shows a low energy cut-off due to the moderate
absorption of \nH$\sim 10^{22}$ \psqcm, which is probably caused by
cold gas in the host galaxy, for example, dust lanes imaged by HST (Malkan et
al 1998), as discussed in Iwasawa et al. (1995). The warm absorber
detected in the Chandra HETGS has the ionization parameter $\xi
=L/(nR^2)$, where $L$ is the ionizing luminosity, $n$ is the density
of the ionized matter, and $R$ is the distance from the ionizing
source, of log $\xi = 2.0$ [erg cm s$^{-1}$], with a column density of
\nH $= 1.5\times 10^{21}$ \psqcm\ (Mocz et al. 2011), which modifies
the spectrum below $\sim 2.3$ keV but has little effect at higher
energies. To avoid unnecessary complication by this warm absorber and
a possible contamination from an extranuclear emission appearing at
lower energies, we limit our spectral study of the nuclear emission
to above 2.3 keV (where a discontinuity of the telescope response due to
Au edges is present). Fitting a simple power-law to the 2.3-11 keV
continuum gives a photon index $\Gamma = 2.08\pm 0.02$ and a
cold
absorption column density of \nH $=(1.3\pm 0.1)\times 10^{22}$ \psqcm,
as measured in the galaxy rest-frame (Table 1). A power-law was
used to describe the continuum spectrum for the analysis below unless
stated otherwise, and this cold absorption was assumed to be
constant. The Fe K emission at 6-7 keV is clearly detected, and there
are possible absorption features on the blue side, as indicated by
arrows in Fig. 1.


\begin{figure}
\centerline{\includegraphics[width=0.45\textwidth,angle=0]{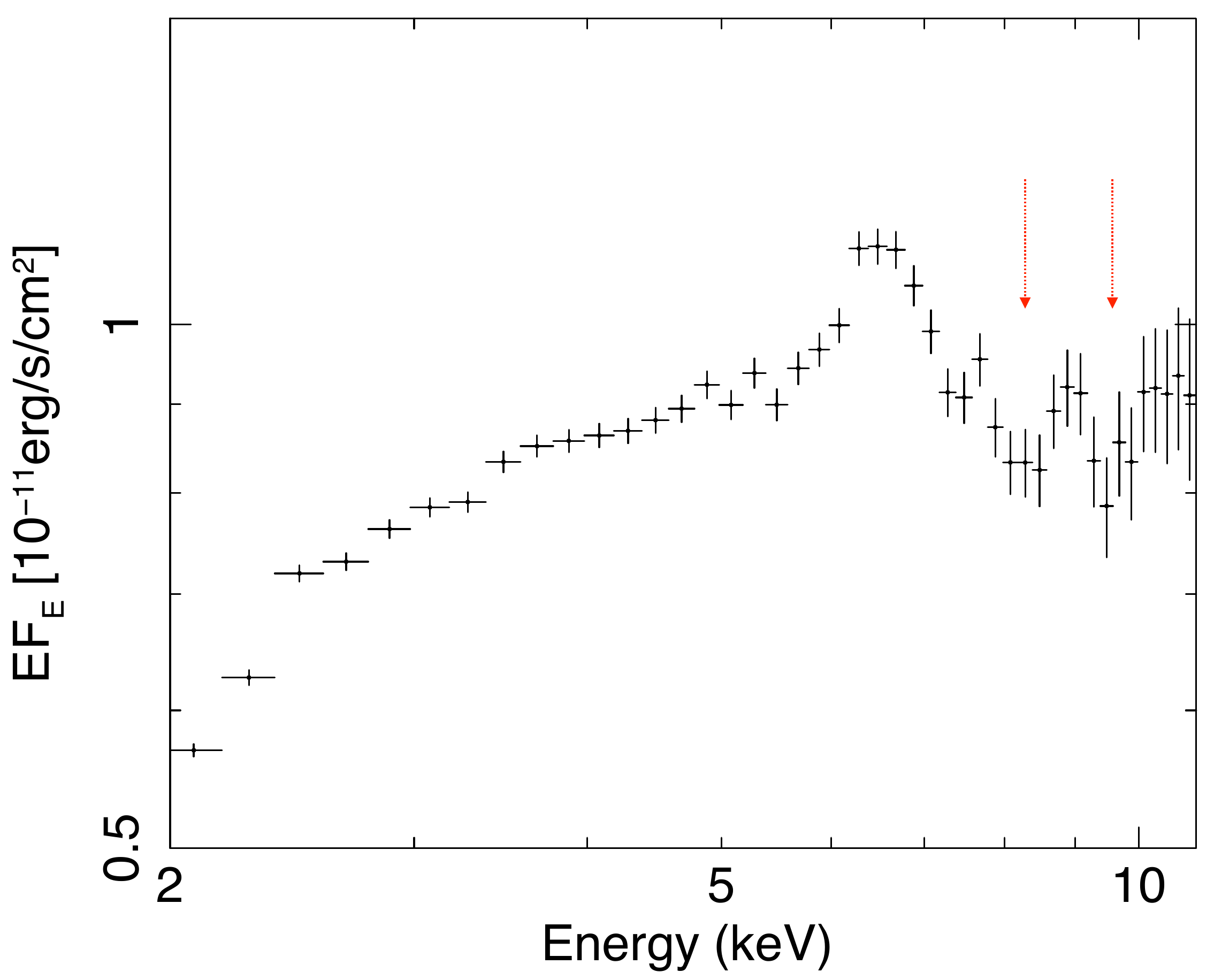}}
\centerline{\includegraphics[width=0.4\textwidth,angle=0]{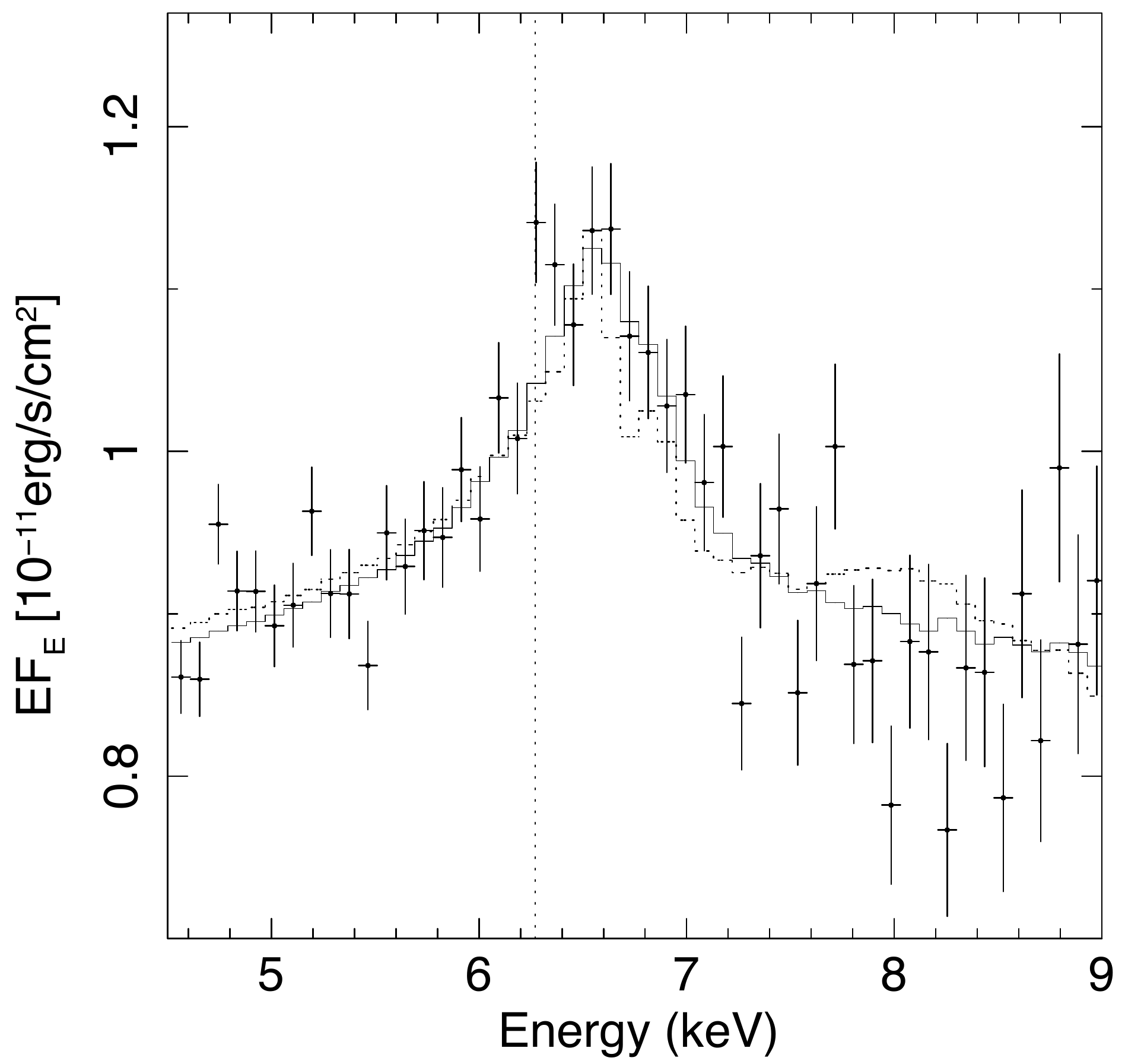}}
\caption{Upper panel: The 2-11 keV spectrum of IRAS 18325--5926
  obtained from the EPIC pn camera. The vertical axis is in flux
  units ($EF_{\rm E}$) and the horizontal axis is the energy (keV) as
  observed. The spectrum is plotted in 200 eV intervals. Two possible
  blueshifted Fe K absorption features are marked by red
  arrows. Lower panel: Details of the Fe-K band spectrum with 90-eV
  energy intervals. The dotted and solid line histograms show the
  best-fit models with {\tt xillver} (log $\xi_{\rm R}\sim 3.5$) and {\tt
    reflionx} (log $\xi_{\rm R}\sim 3.2$), respectively. No relativistic
  blurring is included. The resolved narrow line cores of Fe {\sc xxv}
  and Fe {\sc xxvi} of {\tt xillver} do not agree with the smooth,
  broad profile of the data, whereas the {\tt reflionx} model fits
  well (see text). A narrow line feature is seen at the observed energy
  at the rest-frame 6.4 keV, as indicated by the vertical dotted
  line.}
\end{figure}

The Fe-K band spectrum is shown in Fig. 1b. The line profile is broad
and largely smooth at the EPIC pn resolution $\sim 150$ eV in
  FWHM in the Fe K band at the epoch of observation (XMM-Newton Users
  Handbook 2.13.1). When a Gaussian is fitted, the Fe K emission is found to
have a centroid energy of $6.61\pm 0.04$ keV, a dispersion of $\sigma
= 0.45\pm 0.05$ keV (or FWHM $\approx $48,000 km s$^{-1}$), and
an intensity of $(3.8\pm 0.4)\times 10^{-5}$ \phpspsqcm, as measured in
the rest-frame. This result is robust against continuum modelling
  (see Table 1) and agrees with that from the MOS cameras (see
  Tripathi et al. 2013). The corresponding equivalent width (EW) is
$0.28\pm 0.03$ keV. The line centroid energy is significantly higher than
6.4 keV, which probably indicates that the emission line originates
  in reflection from a highly ionized medium (see Iwasawa et al. 1996
  for a discussion of the hypothesis of cold reflection from an inclined
  disc), but its origin is discussed in detail in Sect. 3.4. In the
  case of reflection from an ionized disc, the main ionization stage
of iron ions would be Fe {\sc xxv} with $\xi $ of the order of $10^3$
erg cm s$^{-1}$. In addition to the broad emission, a small spike is found
at the rest-frame 6.4 keV. Constraints on the energy and width of this
narrow feature need the data of 90 eV binning, as shown in Fig. 1b,
compatible with the spectral resolution. Fitting a narrow Gaussian to
this feature gives a $3 \sigma $ detection with an intensity of $I =
(3.1\pm 1.0)\times 10^{-6}$ \phpspsqcm\ and a corresponding EW of
$\simeq 20$ eV, indicating that the contribution of reflected light from
distant cold matter is minor.

\begin{table*}
\begin{center}
\caption{Mean EPIC pn spectrum of IRAS 18325--5926}
\begin{tabular}{cccccccccc}
Model & $\Gamma $ & $N_{\rm H,A}$ & $E$ & $\sigma$ & $I$ & log $\xi_{\rm O}$ & $N_{\rm H,O}$ & $v/c$ & $\chi^2$/dof \\
(1) & (2) & (3) & (4) & (5) & (6) & (7) & (8) & (9) & (10) \\[5pt]
PG & $2.08\pm 0.02$ & $1.3\pm 0.1$ & $6.61\pm 0.04$ & $0.45\pm 0.05$ & $3.8\pm 0.4$ & --- & --- & --- & 39.1/37\\
PGO & $2.06\pm 0.01$ & 1.3 & $6.61\pm 0.04$ & $0.41\pm 0.05$ & $3.4\pm 0.4$ & $3.9\pm 0.4$ & $5.7\pm 4.4$ & $0.18\pm 0.01$ & 29.0/35 \\
RO$^{\dagger}$ & $2.05\pm 0.02$ & $1.3\pm 0.1$ & --- & --- & --- & $3.9\pm 0.4$ & $5.4\pm 3.8$ & $0.18\pm 0.01$ & 24.8/34 \\
\end{tabular}
\begin{list}{}{}
\item[] Spectral fitting results on the 2.3-11 keV mean spectrum
  obtained from the EPIC pn camera. The spectral data are plotted
in 200 eV
  intervals. (1) Fitted models are PG: power-law modified by
  cold absorption plus a Gaussian line for Fe K; PGO: PG also
modified
   by an outflowing ionized absorber computed by XSTAR; and RO:
  power-law supplemented by reflection from an ionized disc with
  relativistic blurring, computed by {\tt relxill} and modified by the
  outflowing ionized absorption and cold absorption. (2) Photon
  index. (3) Column density of the cold absorption measured in the galaxy
  rest-frame and in units of $10^{22}$\psqcm. (4) Centroid energy of
  the Fe K emission measured in the galaxy rest-frame in units of keV,
  fitted by a Gaussian. (5) Fe K emission line width in units of keV.
  (6) Intensity of the Fe K emission line in units of
  $10^{-5}$\phpspsqcm. (7) Logarithmic value of the ionization
  parameter $\xi_{\rm O}$ of the outflowing absorber in units of erg cm s$^{-1}$. (8)
  Column density of the ionized absorber in units of $10^{22}$\psqcm.
  (9) Blueshift of the absorber in units of light velocity. (10)
  Chi-squared value of the fit and the degrees of freedom. $^\dagger
  $The best-fit parameters of the reflection model are listed in Table
  4 and are discussed in Sect. 3.4. 
\end{list}
\end{center}
\end{table*}


Two possible absorption features are found at $8.43\pm 0.11$ keV and
$9.65\pm 0.12$ keV (as measured in the galaxy rest-frame). They may be
blueshifted high-ionization Fe K absorption lines originating from
high-velocity outflows, as found in a number of Seyfert galaxies and
quasars (e.g. Reeves et al. 2009; Pounds et al. 2006; Chartas et al.
2002, 2003; Tombesi et al. 2010; Vignali et al. 2015), but their
detection is not secure here on statistical grounds. These absorption
features are found to be transient and appear strongly in limited time
intervals, as shown below, where a statistical test supports their
detection. They are therefore diluted in the mean spectrum, leaving
only a weaker trace of their presence. We examine the absorption
features in the mean spectrum first. The same method is applied for
transient absorption features presented in the later section.

The statistical test for the absorption-line feature was adopted from
Tombesi et al. (2010, see also Zoghbi et al. 2014). An absorbed
power-law plus a Gaussian (for the Fe K emission) was used as a
base-line model. The reduction of $\chi^2$ when a Gaussian absorption
line is introduced to explain the spectral drops at 8.4 keV and 9.6
keV is comparable with $\Delta\chi^2=-4.4$ for both features. With
this $\chi^2 $ reduction set to the target $\Delta\chi^2$,
improvements of fit quality were examined for 1000 faked spectra
simulating the base-line model when a narrow Gaussian was introduced
in the 7-10 keV range (because a highly blueshifted Fe K absorption line
can appear anywhere in this energy range, depending on the outflowing
velocity of the absorbing matter). The probability to have a false
line feature exceeding the target $\Delta\chi^2$ was found to be
$P=0.3$. Thus each of the possible absorption features has no
statistical significance. However, the chance probability of two
absorption features with $\Delta\chi^2=-4.4$ at the same time, as seen
in Fig. 1, is deduced to be $P\simeq 0.3\times 0.3 = 0.09$. Detection
of transient absorption features at higher significance are
presented in Sect. 3.3. Given the transient nature and variability of the absorption as described below, the results of the spectral fitting with the XSTAR model for absorption by photoionized gas presented in Table 1 may not reflect the true properties of the absorber. Details of the XSTAR model are described in Sect. 3.3.1.

\subsection{Flux variability}

The mean 2-10 keV flux during the XMM-Newton observation was
$1.4\times 10^{-11}$ \ergpspsqcm, at the lowest level compared to the
other observations of this source (Iwasawa et al. 2004). The
corresponding 2-10 keV luminosity corrected for the internal
absorption (obtained from the absorbed power-law fit to the continuum) is $1.5\times 10^{43}$ \ergps. 


\begin{figure}
\centerline{\includegraphics[width=0.45\textwidth,angle=0]{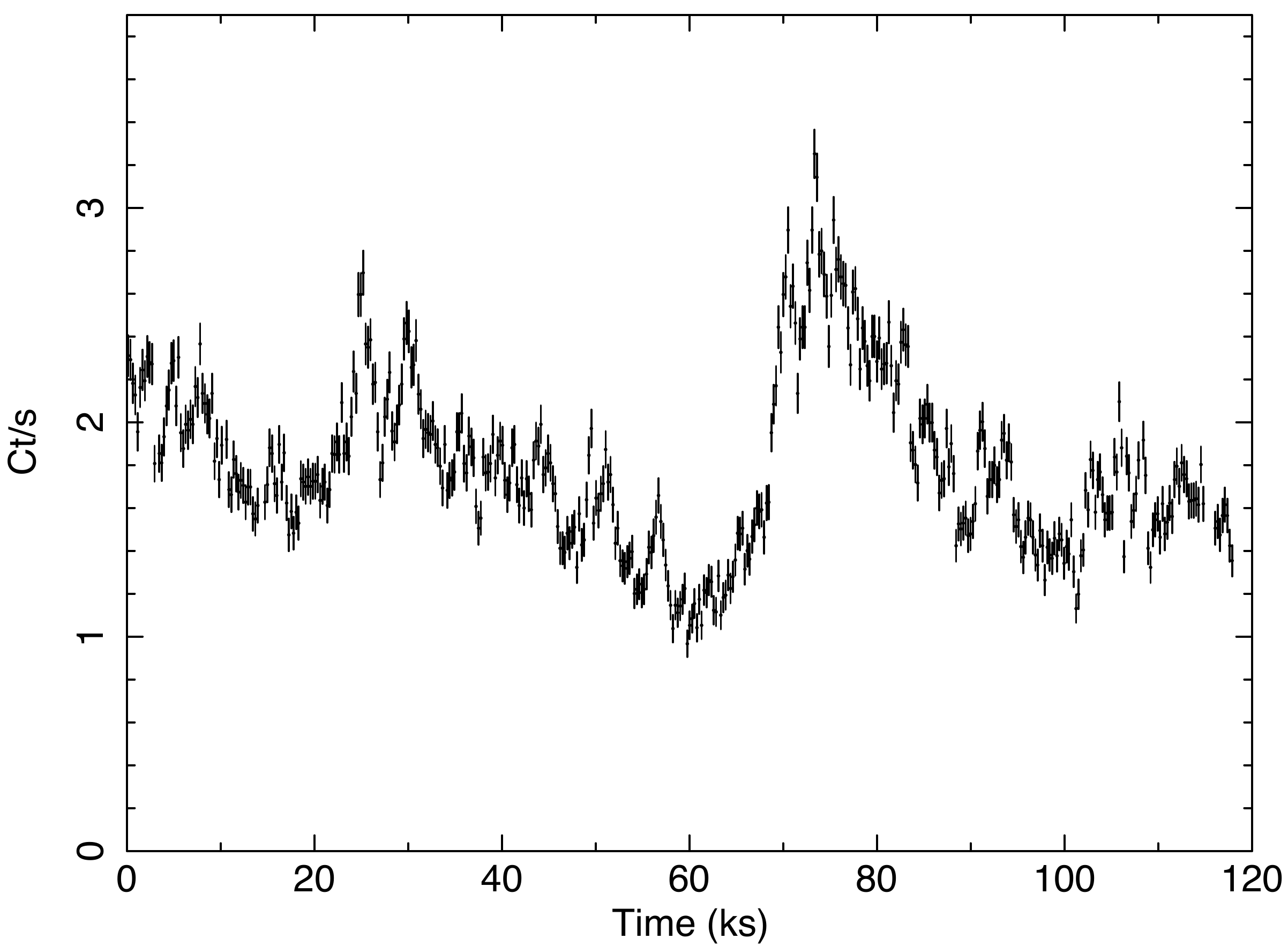}}
\caption{Light curve of IRAS 18325-5926 during the XMM-Newton
  observation obtained from the EPIC pn. The count rates in the 1-8
  keV band corrected for background in 256-s intervals are
  plotted.}
\end{figure}

\subsubsection{Variability amplitude as a function of energy}


The 1-8 keV light curve is shown in Fig. 2. The fractional rms variability
amplitude, $F_{\rm var}$ (Vaughan et al. 2003), is $0.22\pm 0.02$. The
variability is dominated by a strong flare around 70 ks, where the
X-ray flux rises from the lowest flux to the highest of the
observation by a factor of 3 within 15 ks. A weaker broad flare is
also seen around 30 ks. Figure 3 shows $F_{\rm var}$ as a function of
energy. The light curves of 2 ks bins in the five bands corrected for
background were used to compute these $F_{\rm var}$.


\begin{figure}
\centerline{\includegraphics[width=0.45\textwidth,angle=0]{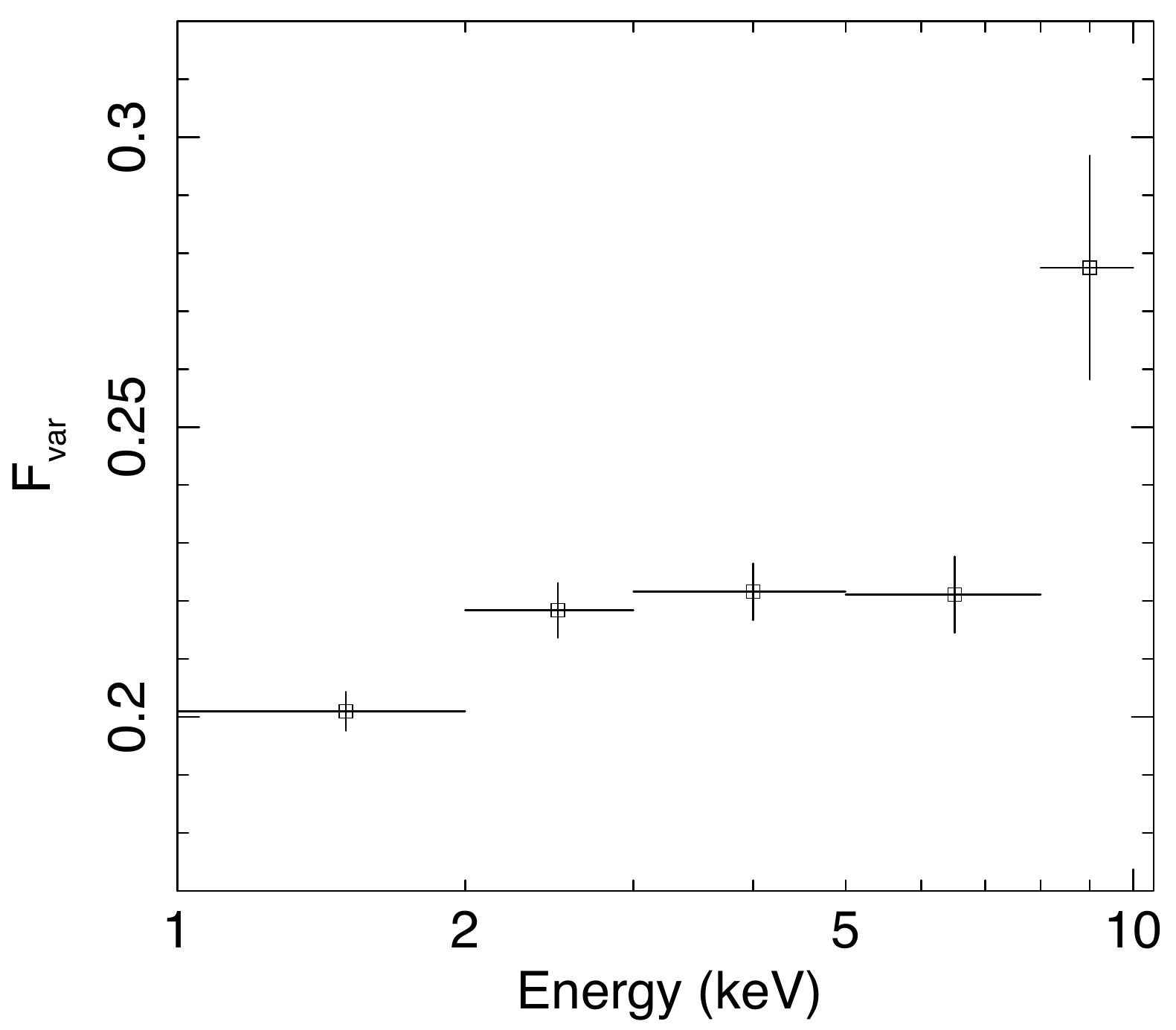}}
\caption{Fractional rms variability amplitude ($F_{\rm var}$, Vaughan et al. 2003) as a
  function of energy is plotted in the 1-10 keV band. They were
  obtained from background-corrected light curves in the respective
  bands for the whole 118 ks duration with 2-ks time bins.}
\end{figure}

$F_{\rm var}$ in the 1-2 keV band drops below the level of the 2-8 keV
band. This decline continues towards lower energies and can be
understood if a contribution of the constant (or weakly variable)
emission from an extranuclear region, which has been discussed in
Iwasawa et al. (1996) and Tripathi et al. (2013), becomes significant
below 2 keV. This justifies the use of the energy range above 2.3 keV
for our spectral analysis of the central source.

Notable is the excess of $F_{\rm var}$ in the 8-10 keV band above that of the
lower energies. This cannot be an artefact caused by incorrect background
correction. The background is, on average, only at a level of 5 per
cent of the 8-10 keV counts. The background light curves taken from
the source-free region is not correlated with the source light
curve, and at the interval of the strongest variation, that is, the flare
around 70 ks, the background is most quiet. A similar excess
of $F_{\rm var}$ in the 8-10 keV band is also seen in the 2013 observation
(Lobban et al. 2014).

\subsubsection{Shape of the light curve as a function of energy}
\begin{figure*}
\centerline{\includegraphics[width=0.75\textwidth,angle=0]{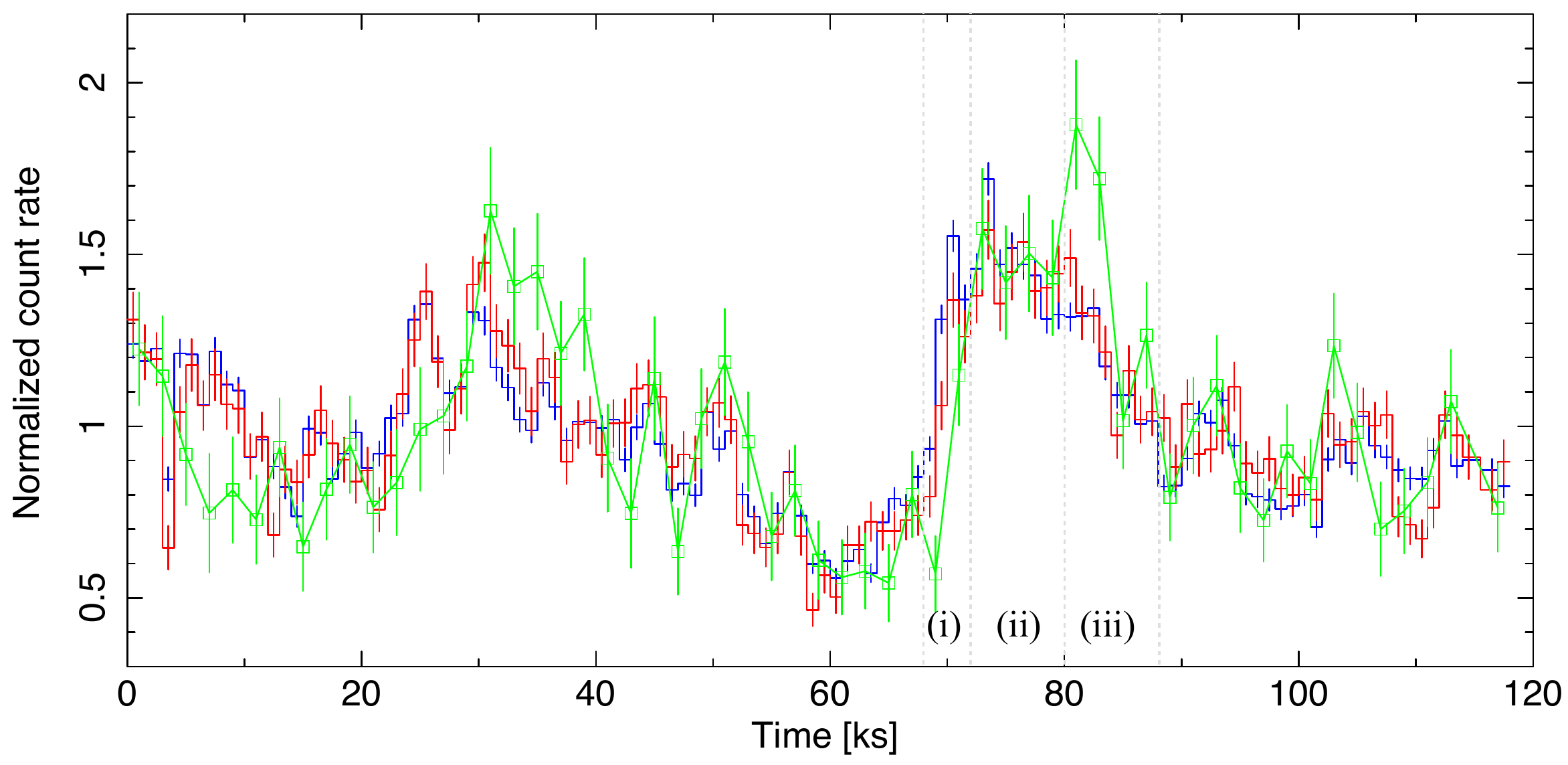}}
\caption{Light curves of the whole observation in 2-5 keV (blue),
  5-8 keV (red), and 8-10 keV (green) bands, normalised to the mean
  count rate in the respective bands. The resolution of the light
  curve is 1 ks for the two lower energy bands and 2 ks for the 8-10
  keV bands. The three intervals during the main flare, (i), (ii), and
  (iii), are indicated (see Sect. 3.3.1, Fig. 6). }
\end{figure*}

The shape of the light curve varies as a function of energy. The
background-corrected light curves in the 2-5 keV, 5-8 keV, and 8-10
keV, normalised by the respective mean count rates, are plotted in
Fig. 4. The normalised light curve apparently shifts behind in time with
increasing energies. This is most notable around the two broad flares at
$\sim 30$ ks and $\sim 70$ ks. The 8-10 keV band, in particular, shows
a clear displacement there from the other lower energy bands. The timing
analysis for the 2013 observation by Lobban et al. (2014) found a hard
lag up to 4 ks, which is qualitatively consistent with the above finding.

Except for the small hard lag, the 2-5 keV and 5-8 keV bands are quite
similar in shape and amplitude. Little or no offset in normalisation
is seen between the two curves. It corresponds that there is little
or no offset in the flux correlation diagram for the two bands. This
means that any constant component with a spectral shape
distinctively different from the variable power-law, such as cold
reflection originating from large radii, as often seen in nearby
Seyfert galaxies, has a minor contribution. If any constant component
were present, it would have a similar spectral shape as the primary
power-law, for example, reflection from high-ionization medium.


The amplitude of the 8-10 keV normalised light curve is larger than
the other two, as also evident in Fig. 3, which causes it to weave
through those of the lower energy bands. The crossing points, where
the 8-10 keV curve rises above those of the lower energies from
beneath, coincide with the peaks of the two broad flares. While the
lagged production of hard X-rays may be accompanied by an amplification of
their flux, the variability of the blueshifted Fe
absorption might, at least partly, account for the excess variability
in the 8-10 keV band, since the absorption uniquely occurs in the
band. This hypothesis with variable absorption and implications from
the behaviour of the 8-10 keV normalised light curve are investigated below.

\subsection{Fe-K band variability}


\begin{figure}
\centerline{\includegraphics[width=0.45\textwidth,angle=0]{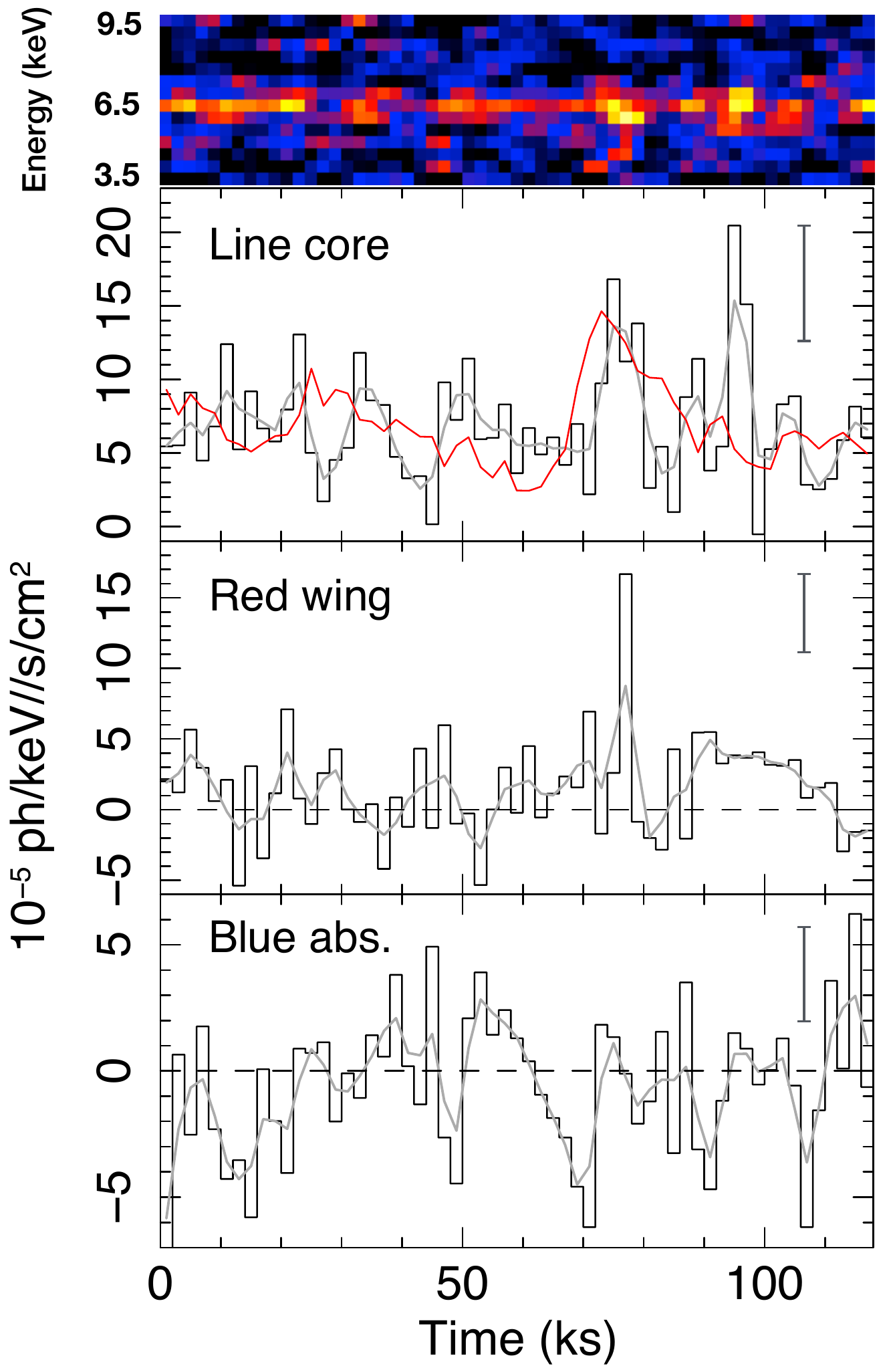}}
\caption{(From top to bottom): The excess map in the time-energy plane
  with a resolution of 2 ks in time and 0.5 keV in energy. Light
  smoothing with an elliptical Gaussian kernel (2 pixel $\times $ 1
  pixel in FWHM) was applied. The three light curves are derived from
  the excess map by projecting the multiple spectral channels,
  corresponding to the Fe K line core centred on the 6.5 keV
  (5.75-7.25 keV) and the red wing (4.25-5.75 keV), and the
  blueshifted absorption band (7.75-9.75 keV). The histograms were
  obtained from the unsmoothed map, while the grey lines stem from the
  smoothed map. The typical measurement error in each band is shown. In
  the panel of the Fe K line core variability, the 2-5 keV continuum
  light curve (the red solid line: adjusted to the mean and variance of
  the line variation) is overplotted for reference. }
\end{figure}

With the X-ray brightness of this source, spectral variability on
short timescales can be studied. To obtain a trend for the variability of
the Fe K spectral features of interest over the whole observation, we
generated an excess emission map, similar to that used in Iwasawa,
Miniutti \& Fabian (2004). It is a map of excess emission (corrected
for the detector response) on the time-energy plane above the
continuum estimated for each time bin. The Fe K emission and
absorption features are relatively broad, and to ensure that each
spectral bin has sufficient counts, we chose a spectral resolution of 0.5
keV and set the time resolution to 2 ks. Each spectrum
has at least 1000 counts in the 2.3-10 keV band even in the faintest
interval. The continuum spectrum was modelled by fitting an absorbed
power-law to the 3.3-10 keV data, excluding the spectral bin of the Fe
emission line peak. The excess map with light smoothing is shown in
Fig. 5. We also show the projected light curves in the three bands, which correspond to
the Fe K line core (5.8-7.3 keV), the red wing (4.3-5.8 keV), and the
blueshifted absorption (8-10 keV), that we extracted from the map. The adjusted continuum light curve is overplotted in red for
reference along the Fe line core light curve.

In general, the data for the Fe-K band features do not show a tight
correlation with the continuum flux. However, there are a few events
that possibly warrant a further investigation. The line core curve
shows a strong rise in flux in two time regions around 76 ks and 96
ks. The first rise over the 8 ks interval roughly coincides with the
main flare, suggesting that the line emission responds to the
continuum. Incidentally, just before the Fe K line flux rise, the blue
absorption band shows a sharp fall of the flux around 70 ks. Since
these notable variations in the Fe K band are associated with the
main flare where the most dramatic flux change in the observation
occurs, we investigate the spectral evolution during this interval
first (Sect. 3.3.1). 

\subsubsection{Rapid evolution of the absorption spectrum during the main flare}

Based on the $\chi^2$ reduction by introducing a Gaussian, the 2 ks
interval spectra around the main flare were examined for a
blueshifted Fe K absorption feature. The two intervals of strongest
suppression in the blue absorption band in the excess map in the 68-72
ks range are verified to have the deepest absorption features with an
equal amount of $\chi^2$ reduction ($\Delta\chi^2=-11$) when a
Gaussian absorption line is introduced. These two intervals are found at similar energies, $8.9\pm 0.2$ keV and $8.6\pm 0.2$ keV, as
measured in the rest frame. This interval is also where the normalised
8-10 keV light curve lies well below those of the lower energy bands
(Fig. 4). Using the simulation method, the chance probability of the
$\chi^2$ reduction in each 2 ks spectrum was found to be
$P=0.015$. Since this occurs over two continuous time-bins in the
whole observation of total 59 bins (which represent the number of
trials), the chance probability is deduced to be $P=0.015\times
0.015\times 59\simeq 0.013$. This value is valid for an absorption line in
each spectrum appearing anywhere in the 7-10 keV. Since the two
intervals show absorptions at energies consistent with each other, the true
chance probability is probably lower than 0.013. The spectrum combining
the two intervals (denoted (i) in Fig. 5: 68-74 ks) is shown in
Fig. 6. A deep absorption feature at $\simeq 8.7$ keV is evident. The
$\chi^2$ reduction when a Gaussian absorption line is introduced is a
$\Delta\chi^2=-21$, which none of 1000 simulations with the baseline
model reaches. 

The absorption spectrum was modelled by the photoionized absorber
computed by XSTAR (Kallman \& Bautista 2001). A relatively broad width
($\sigma = 0.4\pm 0.2$ keV) and a large line EW of $\sim -0.6$ keV
suggest contributions of both Fe {\sc xxv} and Fe {\sc xxvi} with a
high turbulent velocity ($> 1000$ km s$^{-1}$) of the absorber (see
the curve-of-growth analysis presented in Tombesi et al. 2011). Thus we
used an XSTAR table for absorption in photoionized gas with $v_{\rm
  turb}$ = 5000 km s$^{-1}$ (Vignali et al. 2015; Tombesi et al. 2011).
Fitting the XSTAR table gives an ionization parameter
of the absorber of log $\xi_{\rm O} = 3.3\pm 0.2$, a column density of
$N_{\rm H}=1.4^{+0.7}_{-0.5}\times 10^{23}$ cm$^{-2}$, and an outflow
velocity $v=0.23\pm 0.02\thinspace c$. The Fe emission line in this
spectrum is weak without clear broadening ($\sigma < 0.4$ keV, see
Table {\bf 3}). For this short interval, softer X-rays rise earlier,
resulting in a softer continuum spectrum $\Gamma = 2.5\pm 0.2$.

The interval investigated above corresponds to the rising part of the
main flare. The following two intervals of 8 ks each correspond to the
flare peak (ii) 72-80 ks and the declining part (iii) 80-88 ks
(Fig. 4), and their spectra are shown in Fig. 6. Interval (ii)
coincides with the time range of the high Fe line core flux (Fig. 5). The
Fe line flux measured by fitting a Guassian for the three intervals
is shown in Table 3. The strong absorption observed in (i)
  may reduce the disc illumination, which results in the weak line
  flux. During interval (ii), the line flux seems to respond to
  the continuum brightening. Scattering from the outflow observed in
  (i) could add some line flux and shape the broad profile as well. The
Fe emission line in (ii) is strong and broad with a redwards
extension, which moves the Gaussian centroid to a lower energy than
usual (Table 3). The three normalised light curves match in
interval (ii), indicating that the broad-band spectral shape
agrees with that of the mean spectrum and that the absorption
feature in the 8-10 keV band has returned to the average depth. The
8-10 keV band curve in Fig. 4 rises above those of the lower energies
in (iii), where the blue absorption feature disappears (Fig. 6, the
90\% upper limit on the column density is \nH $< 1.2\times 10^{22}$
\psqcm, assuming the same $\xi_{\rm O}$ and blueshift for (i)).

\begin{figure}
\centerline{\includegraphics[width=0.35\textwidth,angle=0]{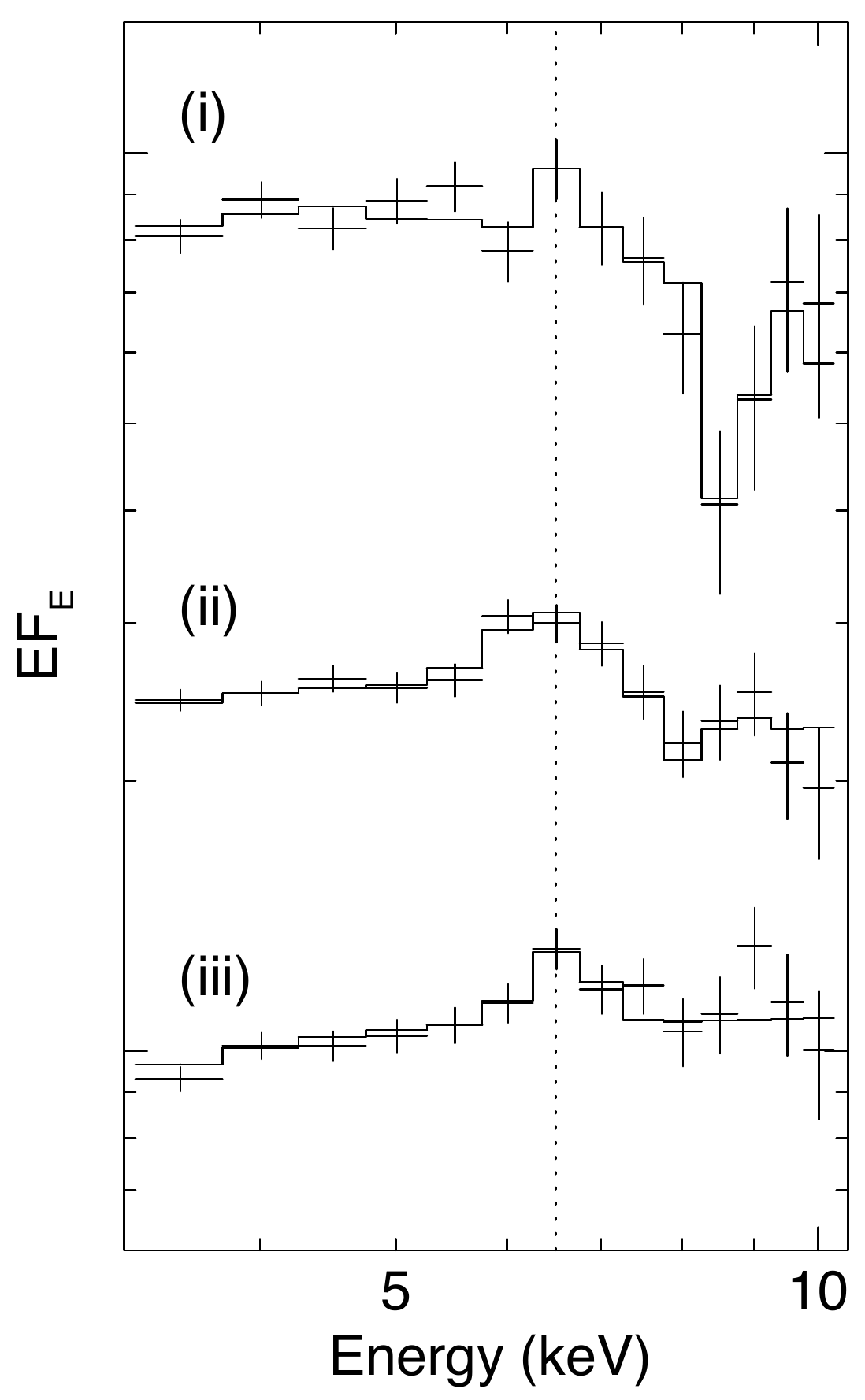}}
\caption{Spectral evolution seen during the large flare. The 3-10 keV
  spectra taken from (i) 68-72 ks, (ii) 72-80 ks, and (iii) 80-88
  ks in Fig. 4 are shown. The histogram indicates the best-fit
  photoionized absorption model (see text). The Fe K emission
    lines are modelled by a Gaussian. The dotted line indicates the
  observed energy of 6.5 keV where the mean Fe K line peaks.}
\end{figure}


\begin{table}
\begin{center}
\caption{Variability of the blueshifted absorber}
\begin{tabular}{ccccc}
Interval & $\Gamma $ & log $\xi_{\rm O}$ & $N_{\rm H,O}$ & $v/c$ \\
(1) & (2) & (3) & (4) & (5) \\[5pt]
(i) & $2.5\pm 0.2$ & $3.3\pm 0.2$ & $14^{+7}_{-5}$ & $0.23\pm 0.02$ \\
(ii) & $2.2\pm 0.1$ & $3.0\pm 0.5$ & $2.1^{+2.0}_{-1.4}$ & $0.17\pm 0.02$ \\
(iii) & $2.0\pm 0.1$ & 3.3 & $<1.2$ & 0.23 \\[5pt]
A & $2.05\pm 0.02$ & $3.8\pm 0.4$ & $4.7^{+12}_{-2.4}$ & $0.19\pm 0.01$ \\
B & $2.03\pm 0.03$ & $3.8$ & $1.0^{+3.0}_{-0.8}$ & $0.08\pm 0.03$ \\
\end{tabular}
\begin{list}{}{}
\item[] Note --- (1) Time intervals where the blueshifted ionized
  absorber is measured by fitting the XSTAR table. (i), (ii), and (iii)
  are intervals during the main flare indicated in Fig. 4, and A and
  B intervals are 0-30 ks and 30-60 ks of the observation,
  respectively. The spectra taken from the A and B intervals are shown
  in Fig. 7. A rise in line core flux occurs, as examined in Sect.
  3.3.3. (2) Photon index of the power-law continuum. (3) Logarithmic
  value of the ionization parameter $\xi_{\rm O}$ of the outflowing absorber in erg cm
  s$^{-1}$. (4) Column density of the absorber in units of $10^{22}$
  cm$^{-2}$. (5) Blueshift of the absorber in units of light
  velocity.
\end{list}
\end{center}
\end{table}


\begin{table}
\begin{center}
\caption{Variability of Fe K emission}
\begin{tabular}{ccccc}
Interval & $E$ & $\sigma$ & $I$ & {\sl EW} \\
(1) & (2) & (3) & (4) & (5) \\[5pt]
(i) & $6.82^{+0.06}_{-0.36}$ & $<0.4$ & $1.9^{+1.1}_{-1.1}$ & 0.13 \\
(ii) & $6.49^{+0.18}_{-0.14}$ & $0.56^{+0.28}_{-0.18}$ & $6.9^{+3.1}_{-2.1}$ & 0.34 \\
(iii) & $6.66^{+0.23}_{-0.15}$ & $0.3^{+0.4}_{-0.3}$ & $2.9^{+2.4}_{-1.2}$ & 0.18 \\[5pt]
94-98 ks & $6.74^{+0.11}_{-0.12}$ & $0.51^{+0.15}_{-0.12}$ & $9.2^{+2.4}_{-2.0}$ & 0.90 \\
\end{tabular}
\begin{list}{}{}
\item[] Note --- (1) Time intervals where the Fe K emission is
  measured by fitting a Gaussian. (i), (ii), and (iii) are intervals
  during the main flare indicated in Fig. 4, and the 94-98 ks interval is
  where the sharp rise in line core flux occurs, as
  examined in Sect. 3.3.3. (2) Gaussian centroid energy in keV,
  measured in the rest frame. (3) Line width represented by Gaussian
  dispersion in keV. (4) Line intensity in units of $10^{-5}$
  \phpspsqcm. (5) Line equivalent width in keV. The narrow line in interval (i) was investigated using the spectrum with a 200-eV resolution.
\end{list}
\end{center}
\end{table}

\subsubsection{Repeated absorption spectrum evolution?}


\begin{figure}
\hbox{{\includegraphics[width=0.25\textwidth,angle=0]{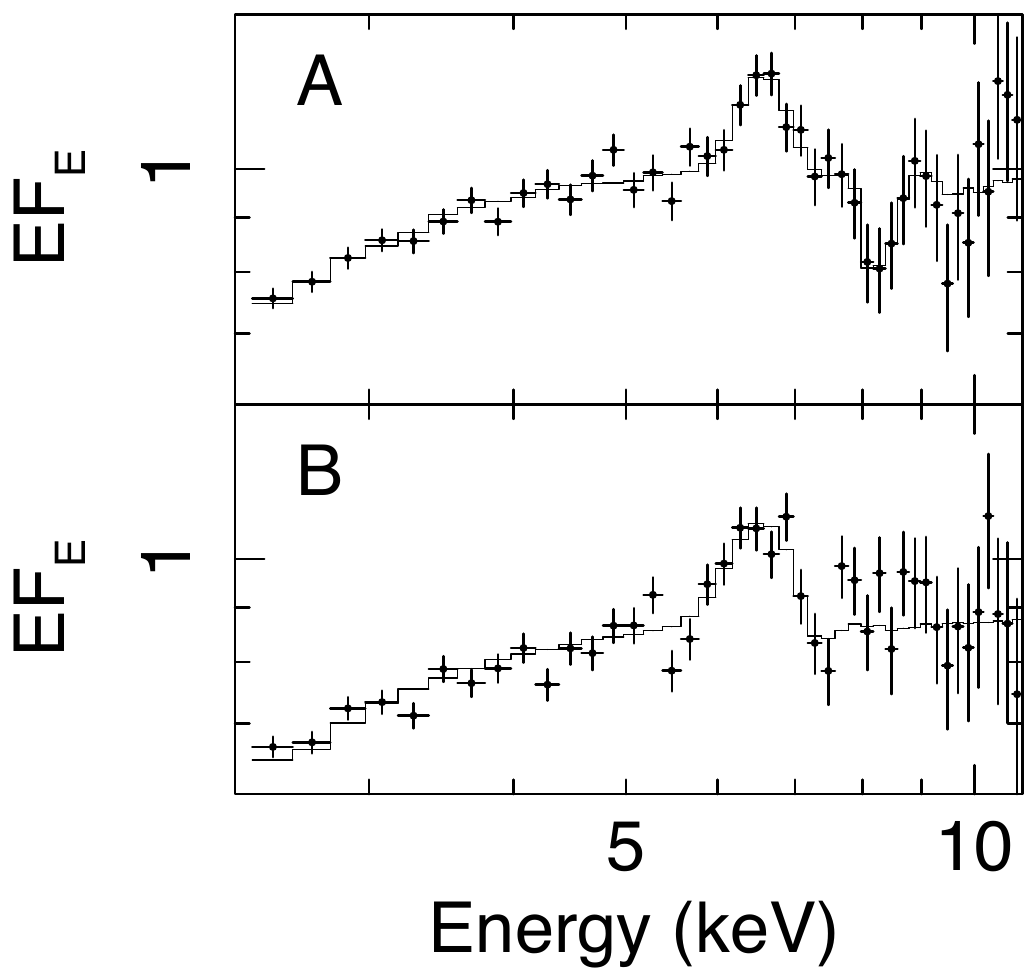}}\vspace{-1mm}
{\includegraphics[width=0.235\textwidth,angle=0]{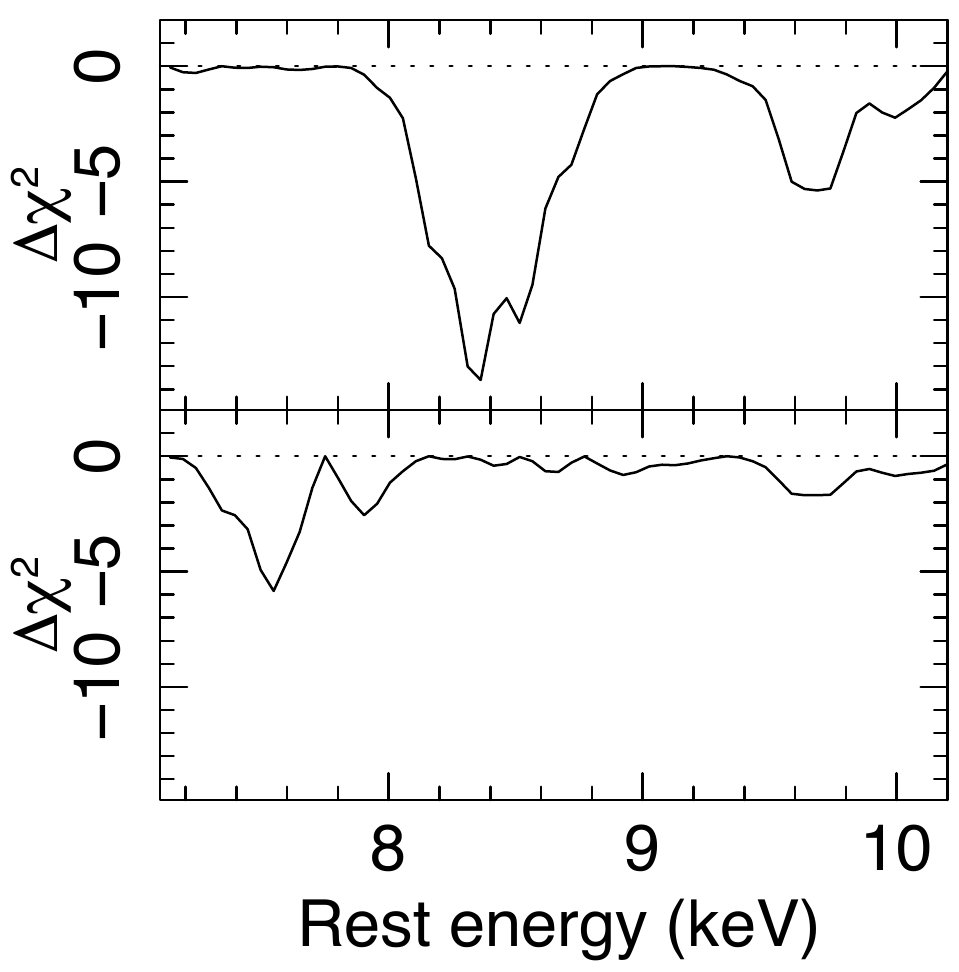}}}
\caption{Left: The 2.3-11 keV spectra taken from the two intervals, A
  0-30 ks and B 30-60 ks, in the light curve. The spectra are plotted
  in 200-eV intervals with the best-fit model of the blueshifted ionized
  absorption computed by XSTAR (see text). Right: The $\chi^2$ curve
  as a function of energy, when scanning with a narrow Gaussian for
  absorption lines in the 7-10 keV range for the corresponding spectra.}
\end{figure}

The dramatic spectral evolution in the blue absorption band occurs
within 30 ks around the main flare. These absorption variations can be
traced by the behaviour of the 8-10 keV normalised light curve
relative to that of 5-8 keV (Fig. 4). Guided by Fig. 4, we assumed
that a similar spectral evolution might be taking place across the
other flare in the first half (0-60 ks) of the observation, albeit in a
less dramatic fashion over a longer time range. Two time intervals
with contrasting 8-10 keV behaviours were investigated.

Figure 7 shows the spectra taken from the intervals of 0-30 ks (A) and
30-60 ks (B). As above, absorption features were searched by scanning
the 7-10 keV by a narrow Gaussian, and the resulting $\chi^2$ reduction is
plotted in the right panel for each spectrum.  The largest reduction
in $\chi^2$ is found at 8.3 keV in the interval A spectrum with
$\Delta\chi^2=-13$. The statistical test finds the chance probability
for this feature to be $P\simeq 0.018$ (including the number of trials
of 3). Furthermore, another absorption feature with
$\Delta\chi^2=-5.5$ is seen at 9.7 keV in the interval A
spectrum. None of 1000 simulated spectra show two line features,
regardless of energy, at the same time with $\chi^2$ reductions
comparable or larger than those two observed absorption features,
suggesting a chance probability of $P<0.003$. Therefore we consider
a detection of the absorption features in interval A as
likely. Fitting the XSTAR table to the stronger 8.3 keV feature gives
log $\xi_{\rm O}=3.7^{+0.6}_{-0.3}$, $N_{\rm H}=(3.8\pm 1.5)\times 10^{22}$ cm$^{-2}$,
and $v=0.19\pm 0.01\thinspace c$. The weaker 9.7 keV absorption
feature coincides with the K$\beta $ transitions, but the observed
depth is larger than expected. If it originates instead from a
higher velocity system, its blueshift would be $v=0.30\pm
0.01\thinspace c$, assuming the same ionization parameter for the 8.3
keV absorption. On the other hand, the spectrum of interval B shows no
clear sign of the absorption feature seen in the interval A spectrum. Interval B is the post-flare interval, as the (iii) interval of
Fig. 6, where the absorption feature disappears.

\subsubsection{Sharp increase in the Fe K line flux}

\begin{figure}
\centerline{\includegraphics[width=0.5\textwidth,angle=0]{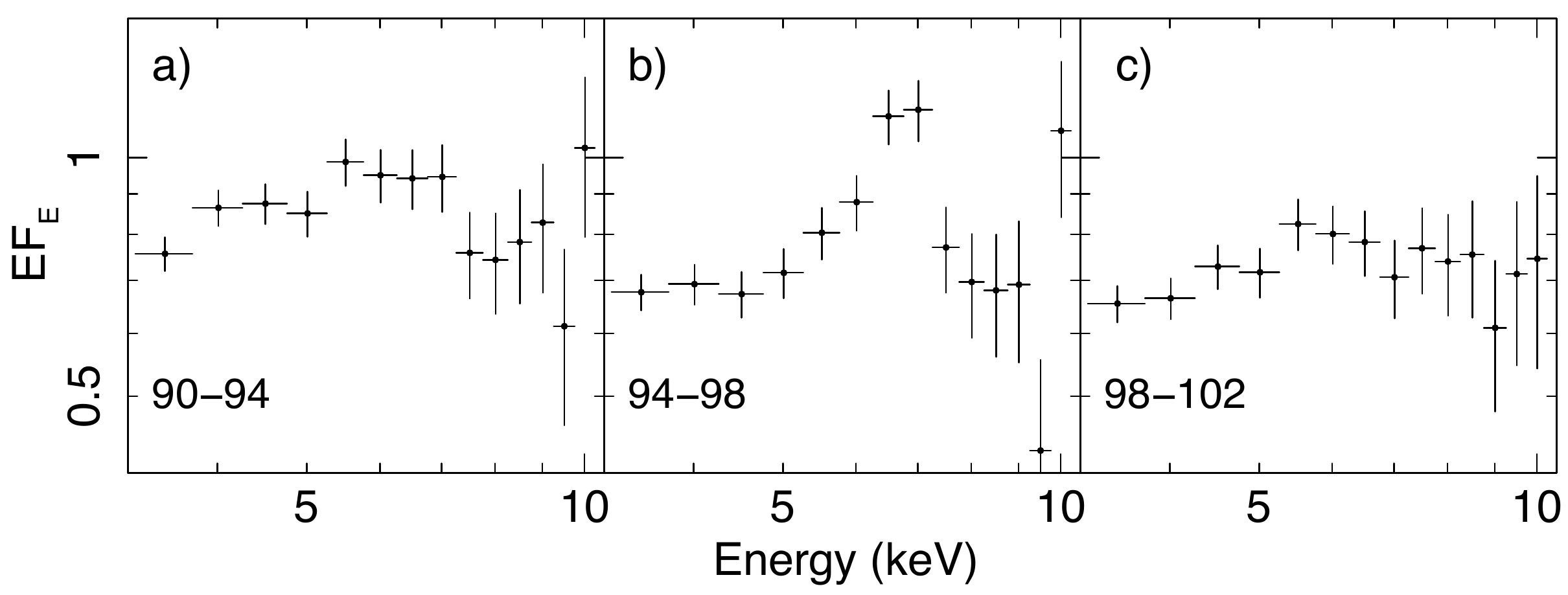}}
\caption{Spectrum of the interval of a short burst of the Fe K
  line core flux (94-98 ks, see Fig. 5, Sect. 3.3.3) is shown in
panel  b). The spectra of the neighbouring intervals with the same 4 ks
  durations are shown in panels a) and c).}
\end{figure}

The 94-98 ks interval shows a sharp rise in Fe K emission flux
(Fig. 5) that is not accompanied by any obvious continuum flux
increase. The two consecutive time bins of 2 ks intervals show
exceptionally strong line flux in the core band, even stronger than
those during the major flare (see Table 3). This band is where the
Fe K emission is expected, therefore we examined how statistically reliable the
high intensities seen in those intervals are, again using simulated
spectra with Fe K emission of the mean spectrum. The line core flux
was derived for each simulated spectrum following the same procedure
to generate the excess map and the light curve in Fig. 5. The
chance probabilities for the fluxes of  94-96 ks and 96-98 ks are $<0.001$ and $0.01$, respectively. These two intervals
of strong intensity occur in a row in the whole observations of 59
time intervals, which implies the chance probability to $P< 6\times
10^{-4}$. A detection of the Fe line flux increase in this short
interval is likely. Figure 5 suggests that it is also accompanied by a flux
increase in the red wing.

Figure 8 shows the spectrum of the 94-98 ks interval along with the
preceding and following intervals of 4 ks. The strong line core in
the 94-98 ks spectrum is evident in contrast to the spectra of the
neighbouring intervals, although broad emission down to 5 keV may be
seen in both panels. The line flux of the 94-98 ks interval obtained by
fitting a Gaussian is $(9.2\pm 2.2)\times 10^{-5}$ \phpspsqcm, is
factor of $\sim 3$ larger than the mean value (Table 1). The continuum
level comparable to the mean flux leads to an extremely large EW of
$0.90\pm 0.21$ keV. This large EW is a result of the increase in absolute
line flux. A Fe K line this strong in absence of
a continuum increase is difficult to understand unless the illuminating
source temporarily becomes anisotropic. Gravitational light focusing
effect (e.g. Martocchia \& Matt 1996; Miniutti \& Fabian 2004) is not
compatible with the line shape because it does not match the
expected extreme
gravitational redshift. Anisotropy needs to be an intrinsic
property of the illuminating source. We note that although not as
extreme as this, unusually strong Fe K emission (EW$\sim 0.6$ keV) has
occasionally been observed in previous ASCA observations (Iwasawa
et al. 2004), suggesting that similar bursts of Fe K line emission
occur routinely in this source.

\subsubsection{Broad emission component}

The red wing band flux fluctuates around null in the first half of the
observation but seems to show a systematic excess in the second half
(Fig. 5). When a Gaussian is fitted to the Fe K line profile of the
spectra taken from the two halves, the line widths are $\sigma =
0.36\pm 0.05$ keV in the first half and $\sigma = 0.60\pm 0.09$ keV in
the second half, respectively, while the line centroid energies are compatible ($6.62\pm 0.04$ keV to $6.54\pm
0.07$ keV, respectively). The enhanced line broadening can come
either from increased $\xi_{\rm R}$ (due to Compton broadening) or stronger
gravitational effects. Given that the mean source luminosity is
comparable between the two intervals, increased relativistic effect is
a more likely cause. This suggests that the line-emitting region
has moved inwards in the latter half of the observation, which is
examined further with the relativistic blurring model below.

\subsection{Modelling the reflection spectrum}

As shown above, the Fe-K band spectrum of IRAS 18325--5926 is
  variable. The blueshifted absorption features appear in limited time
  intervals, indicating that they are associated with transient
  outflows. The Fe K emission line is, in contrast, a persistent
  feature throughout the observation (Fig. 5) and maintains a similar shape apart from some intervals of dramatic variability examined in Sect. 3.3. This favours a hypothesis that the line originates primarily in reflection from the accretion disc and
not from the transient outflow, which could also produce line emission. While the transient
  absorption features are highly diluted, the emission line in the
  mean spectrum represents the average properties of the disc
  reflection as the strong variability mentioned above is seen in limited time intervals. Although the broad emission-line feature can be composed of blended multiple line components, the smooth profile at the resolution of the pn camera and the observed variability argue for reflection from the accretion disc as the origin of the Fe K emission.

Since the Gaussian modelling of the Fe K line suggests reflection from
highly ionized matter (Sect. 3.1), we tried to model the mean
spectrum with an ionized reflection spectrum in addition to the
primary power-law. While the Fe K feature is clearly broad, part of
the broadening has to be due to Compton scattering if the line
originate from highly ionized medium since line photons travel through a
hot medium before escaping from it. This Compton broadening depends on
models to some degree. The line profile computed by {\tt reflionx}
(Ross \& Fabian 2005), which has been a favourite choice of many,
exhibits a symmetrical, classic broad shape and fits the data well
with log $\xi_{\rm R}\simeq 3.2$ without any extra broadening
(Fig. 1b). However, here we used {\tt xillver} (Garc\'ia et al. 2013)
instead because of advantages on physical grounds, particularly
the use of a richer set of updated atomic data base and the
availability of the angle-dependent spectra, as opposed to the
angle-averaged spectra of {\tt reflionx}. At log $\xi_{\rm R}\sim 3$, the line
shape predicted by {\tt xillver} is composed of narrow cores of Fe
{\sc xxv} and Fe {\sc xxvi}, which presumably come from a thin, upper
layer of the ionized surface (they resemble those in the
  reflection spectrum of {\tt xion} (Nayakshin et al. 2000) for an
  ionized disc in a hydrostatic equilibrium. The {\tt xion} model was
  used in Iwasawa et al. (2004), in which relativistic broadening was
  required to explain the observed line profiles because of its
  intrinsically narrow Fe K feature) on a Compton-broadened
base. The temperature of the disc is generally found lower, that
is, less
Compton broadening, in {\tt xillver} than in {\tt reflionx}. A
detailed comparison between the two models can be found in Garc\'ia et
al (2013). As shown in Fig. 1b, the narrow line cores of {\tt
  xillver}, which would be resolved at the resolution of the pn
camera, are not seen in the observed profile. The use of {\tt xillver}
thus requires, unlike in {\tt reflionx}, an extra broadening
mechanism, such as relativistic broadening, which smears them out.

The {\tt xillver} reflection spectrum table is integrated with the
relativistic blurring kernel {\tt relline} (Dauser et al. 2010) and the
combined model {\tt relxill} (Garc\'ia et al. 2014) is avialable. The
2.3-11 keV mean spectrum was fitted by the relativistically blurred
reflection model added to an illuminating power-law. The cold
absorption and the blueshifted ionized absorption were fitted. The
fitting result is given in Table 1 (for the power-law continuum and
absorption) and Table 4 (for the reflection). The ionization parameter
of the reflection spectrum is found to be log $\xi_{\rm R} = 3.0\pm 0.2$. The
reflection fraction is moderate with $R\sim 0.57$, which translates
into $\sim 20$\% of the 2-10 keV flux coming from the reflection
spectrum. Our view to the disc is inferred to be intermediate
($i=42^{\circ }\pm 4^{\circ }$). Given the moderate line broadening,
the required relativistic effect is accordingly moderate and the black
hole spin is unconstrained. All the negative values, corresponding to
the retrograde cases with the innermost stable orbit (ISCO) up to
$9\thinspace r_{\rm g}$ (Novikov \& Thorne 1973), are permitted. The
small disc radii near the ISCO in this source are, on average, not
well illuminated because of the source height (Fabian et al. 2014) or the
inner disc configuration as discussed later (Sect. 4.1).

Since the Fe K line width changes between the first and second halves
of the observation (Sect. 3.3.4), which suggests a change in
relativistic effect, we tried to see how it is reflected in the
reflection model by comparing the two intervals. We let the
emissivity index (of the inner radii) free in the fits, while the other
relativistic blurring parameters were fixed to those obtained for the mean
spectrum, including the unconstrained spin parameter set to be
$a=0.1$. The emissivity index is inherently not a well-constrained
parameter, as seen in Table 4, but, by comparing the best-fit values,
$0.1(<1.9)$ for the 0-60 ks interval and $1.5(<2.4)$ for the 60-118 ks
interval, the line emitting region in the second interval might be
more concentrated in the smaller radii than in the first, in agreement
with the line width change. Constraints of the emissivity index remain loose when a higher black hole spin (which is unconstrained) is assumed.

\begin{table}
\begin{center}
\caption{Spectral fits with the reflection model}
\begin{tabular}{llccc}
Data && Total & 0-60 ks & 60-118 ks \\[5pt]
Continuum & $\Gamma $ & $2.05\pm 0.04$ & $2.00\pm 0.03$ & $2.05\pm 0.03$ \\
& $N_{\rm H,C}$ & $1.3\pm 0.1$ & 1.3 & 1.3 \\[5pt]
Reflection & log $\xi_{\rm R}$ & $3.0\pm 0.2$ & $2.9\pm 0.2$ & $3.1\pm 0.2$ \\
 & $R$ & $0.57\pm 0.08$ & $0.44\pm 0.09$ & $0.56\pm 0.13$\\[5pt]
Blurring & $i$ & $42\pm 4$ & 42 & 42 \\
 & $a$ & $0.1^{\dagger}$ & 0.1 & 0.1 \\
 & $\beta $ & 2 & 0.1 ($<1.9$) & 1.5 ($<2.4$) \\[5pt]
\end{tabular}
\begin{list}{}{}
\item[] Note --- The spectral data obtained for the whole duration
  (Total), first (0-60 ks) and second (60-118 ks) halves of the
  observation were fitted by the reflection model. All the three
  spectra are of 200-eV intervals in the 2.3-11 keV band. The fitted
  model is a power-law supplemented by reflection from an ionized disc
  including relativistic blurring, computed by {\tt relxill} (Garc\'ia
  et al. 2014). Cold and ionized blueshifted absorption are also
  included. The $N_{\rm H}$ of cold absorption was fitted for the
  total spectrum and fixed for the other two. The best-fit
  parameters of the ionized absorption are comparable among the
  three (see Table 1). The ionization parameter of the reflecting
  medium is $\xi_{\rm R}$. $R$ is the reflection fraction with respect to
  that from $2 \pi$ solid angle. In modelling relativistic
  blurring, $i$ is the inclination angle in degree, $a$ is the
  dimensionless parameter of the black hole spin, and $\beta $ is the
  slope of radial emissivity law ($\propto r^{-\beta}$). The inner
  disc radius is automatically set by the spin parameter $a,$ while the
  outer radius is assumed to be $1000\thinspace r_{\rm g}$. For the
  0-60 ks and 60-118 ks spectra, the emissivity law is assumed to have
  a break at $100\thinspace r_{\rm g}$ and the inner $\beta$ is fitted, while the outer $\beta$ is fixed at 2. $\dagger $: The black hole
  spin $a$ is unconstrained. The permitted range includes negative
  values (for a retrograde disc) down to $-0.998$.
\end{list}
\end{center}
\end{table}

\section{Discussion}

\subsection{Black hole mass and accretion rate}

Among nearby Seyfert galaxies, the reflection signature from the highly
ionized medium in IRAS 18325--5926 is a rare example, possibly
indicating an extreme condition of accretion flow in the active
nucleus.

No optical reverberation measurement of $M_{\rm BH}$ is available for
this Seyfert 2 galaxy. An alternative method for estimating $M_{\rm BH}$
is to use the $M_{\rm BH}$-$\sigma_{\star}$ relation by using a [O{\sc
  iii}] $\lambda 5007$ line width as a proxy to stellar dispersion
$\sigma_{\star}$. Iwasawa et al. (1995) decomposed the [O{\sc iii}] line
profile into a blueshifted broad component and a
narrow component. The FWHM of the narrow component is 450 km s$^{-1}$
or $\sigma = 190$ km s$^{-1}$, and the other narrow emission
lines
show similar line widths. Although the calibration between [O{\sc
  iii}] and stellar dispersion differs in the samples (see Onken et
al 2004), $M_{\rm BH}$ of IRAS 18325--5926 is always found to be
around $~10^8 M_{\odot}$. Dasyra et al. (2011) used narrow-line region
(NLR) properties in near-IR obtained from Spitzer and estimated
$M_{\rm BH}$ to be $1.7\times 10^8\thinspace
M_{\odot}$ or $7.0\times 10^8\thinspace M_{\odot}$ depending on the
estimating method. The rapid X-ray variability seems incompatible with
this high black hole mass. As warned by many, the use of [O{\sc iii}]
data as a substitute for $\sigma_{\star}$ can fail dramatically in
estimating $M_{\rm BH}$ of an individual object (e.g. Onken et al.
2004). The presence of the blueshifted component of [O{\sc iii}]
suggests that an outflow probably affects the NLR kinematics, and the
high-velocity outflow caught in the X-ray observation in the nucleus
may have an influence at larger radii. 

A good correlation between X-ray variability excess variance
($\sigma_{\rm rms}^2$) and the black hole mass ($M_{\rm BH}$) was
found for a sample of nearby bright (unobscured) Seyfert galaxies
(Ponti et al. 2012). We obtained the excess variance for IRAS
18325--5926, using the 2-10 keV light curves with a 500 s resolution
as done in Ponti et al. (2012). Three 80 ks portions were taken from
the whole light curve, and by sliding the starting time by 20 ks, excess
variances for the three light curves were computed and their mean was
taken as the representative value, $\sigma_{\rm rms}^2=(6.3\pm
0.7)\times 10^{-2}$. Using their relation for 80 ks light curves,
calibrated by reverberation masses, the $M_{\rm BH}$ of IRAS 18325--5926
is found to be $2.3\times 10^{6}M_{\odot}$. The scatter of this
correlation is estimated to be 0.44 dex (or a factor of 3). 

The spectral energy distribution of this IRAS-selected galaxy is
peaked at the infrared with a luminosity of $L_{\rm ir}=10^{11.1}L_{\odot}$,
or $4.8\times 10^{44}$ \ergps\ (Kewley et al. 2001).
Given the warm IRAS colour, the infrared emission is probably dominated
by dust reradiation of the obscured active nucleus (de Grijp et al.
1985). This is supported by the high AGN index $D_{\rm AGN}=0.8$ (which does not represents an actual AGN fraction in bolometric
  luminosity), based on the
optical spectroscopy (Yuan, Kewley \& Sanders 2010). Using the
historic mean of the X-ray luminosity of IRAS 18325--5926
($L_{2-10}\sim 2\times 10^{43}$ \ergps, see e.g. Iwasawa et al. 2004)
and the bolometric correction ($L_{\rm bol}/L_{2-10} \simeq 20$)
derived in Marconi et al. (2004), the AGN bolometric luminosity is
estimated to be $L_{\rm bol}\sim 4\times 10^{44}$ \ergps, which
strongly suggests AGN dominance of the bolometric luminosity. It also
means that when the black hole mass inferred from the X-ray
variability is adopted (the classical Eddington luminosiy for this
$M_{\rm BH}$ is $L_{\rm Edd}\simeq 3.0\times 10^{44}$ \ergps), the AGN
luminosity is comparable to the Eddington luminosity and the black
hole may be operating at critical (= Eddington) or supercritical
accretion. This Eddington ratio may have a large uncertainty that is mainly attributed to the black hole mass
  measurement. However, a supercritical or critical accretion onto a small black
  hole explains more naturally the rapid-variability highly
  ionized disc (e.g. Ross \& Fabian 1993) and strong outflow (e.g.
  Ohsuga et al. 2009) observed in IRAS 18325--5926 than otherwise.

\subsection{Ionized disc and transient high-velocity outflows}

Strong radiation produced at the high accretion rate causes a highly
ionized disc surface, as indicated by the Fe K
emission. Close to the critical accretion, the thermal effect inflates
the innermost radii of the disc, which may impede full illumination of
the inner radii of the accretion disc by the X-ray source. This is
compatible with the moderate reflection fraction ($R\sim 0.6$) and
the moderate relativistic broadening of the Fe K line. If the inner
disc radii close to the ISCO is not fully illuminated, the black hole
spin parameter obtained from the Fe K line shape (Table {\bf 4}) would not
reflect the real spin (as discussed in detail by Fabian et al.
2014). Nonetheless, the response of the Fe emission line to the major
flare of the continuum (Fig. 5, Fig. 6) supports the Fe K line
emission to be of the disc origin. Since the Compton-broadened
component resulted from multiple scatterings in the disc atmosphere,
variability of this component should be diluted, meaning that the response of
the total line flux to the continuum is expected to be slow to some
degree. On the other hand, the brief brightening of the Fe line in the
absence of a continuum flux increase, as shown in Sect. 3.3.3, is
difficult to explain in the context of disc reflection.

Strong outflows are also expected from the central part of the
accretion disc (e.g. King et al. 2004; Ohsuga et al. 2009), with an
increasing outflow rate towards small radii (Takeuchi, Mineshige \&
Ohsuga 2009). High-velocity Fe absorption features detected in the
X-ray spectra of IRAS 18325--5926 might be identified with such
outflows driven by radiation pressure, for which the outflow velocity
could reach $\sim 0.2\thinspace c$ (e.g. Ohsuga et al.
2009; Proga \& Kallman 2004; Sim et al. 2010).

As shown in Sects. 3.3.1 and 3.3.2, the Fe absorption feature in IRAS
18325--5926 is transient. The deepest absorption shown in Fig. 6 lasts
for 4 ks. Variability of Fe K absorption features between observations
separated by days to years has been often found in well-studied
objects, for instance, Chartas et al. (2009) and Tombesi et al. (2011). Faster
variability within a single observation was reported for BAL QSO PG
1126--041 by Giustini et al. (2011). The variability seen in IRAS
18325--5926 is one of the rapid extremes. It could be explained
in part by the mass scaling that is due to the light black hole, but it also may
reflect the turbulent nature of the flow. 

An interesting question is whether this outflow variability has any
connection with the disc activity. The deep absorption was observed at
the onset of the main flare. Incidentally, the other likely detection
of the absorption feature in the interval A spectrum (0-30 ks,
Fig. 7) precedes the flare at around 30 ks. Whether these two events are
chance coincidence needs to be tested by further observations. The
loss of angular momentum by outflows induces an increased accretion
that leads to a disc flare, but the viscous timescale is too long to
match the delay of the continuum flare, which is instead more
comparable with the dynamical timescale of $\approx 2.3
(r/10\thinspace r_{\rm g})^{3/2}M_{6.3}$ ks for the X-ray variability
mass $M_{6.3}$. In both incidences, the absorption feature disappears
after each flare peak: (iii) of Fig. 6, and interval B of Fig. 7,
repectively, resulting in similar spectral evolution across the two
flares. If this disappearance is caused by overionization and
not by the outflowing matter going past our view, the lowered density of the
outflow is more likely than the increased illumination because the
required increase of illuminating luminosity is short of more than an
order of magnitude. Rapid thermal expansion of the outflowing matter, caused by heating during the flare, for example, may cause 
the disappearance of the absorption feature after the peak of the flare (although we cannot track the evolution of the ionization with the current data quality). 
The sequence of spectral evolution around the
flares in the Fe absorption feature and the continuum showing a
hard lag are phenomenologically similar to the feature seen in GRS 1915+105 in
the heartbeat state (Neilsen, Remillard \& Lee 2011), although the
timescale is $\sim 2$ orders of magnitude shorter in the IRAS galaxy
when scaling with $M_{\rm BH}$.

\begin{acknowledgements}
KI acknowledges support by the Spanish MINECO under grant 
AYA2013-47447-C3-2-P and MDM-2014-0369 of ICCUB (Unidad de 
Excelencia 'Mar\'ia de Maeztu'). 
\end{acknowledgements}

\end{document}